\documentclass[aps,prev,twocolumn,preprintnumbers,floatfix,nofootinbib]{revtex4-1}
\usepackage{graphicx}
\usepackage{bm}
\usepackage{times}
\usepackage{slashed}
\usepackage{color}
\usepackage{appendix}
\usepackage{mathtools}
\usepackage{epsfig}
\usepackage{dcolumn}
\usepackage{textcomp}
\newcommand{\red}{\color{red}}
\newcommand{\beq}{\begin{equation}}
\newcommand{\eeq}{\end{equation}}

\pdfoutput=1

\begin{document}

\preprint{IIPDM-2020}

\title{XENON1T Anomaly: A Light $Z^\prime$ from a Two Higgs Doublet Model}

\author{Manfred Lindner $^{a}$}\email{lindner@mpi-hd.mpg.de}
\author{Yann Mambrini $^{b}$}\email{yann.mambrini@th.u-psud.fr}
\author{T\'essio B. de Melo$^{c}$}\email{tessiomelo@gmail.com}
\author{Farinaldo S. Queiroz$^{c}$}\email{farinaldo.queiroz@iip.ufrn.br}

\affiliation{$^a$ Max Planck Institut f\"ur Kernphysik, Saupfercheckweg 1, 69117 Heidelberg, Germany\\
$^c$ Universit\'e Paris-Saclay, CNRS/IN2P3, IJCLab, 91405 Orsay, France\\
$^c$~International Institute of Physics,
Universidade Federal do Rio Grande do Norte, Campus Universit\'ario, Lagoa Nova, Natal-RN 59078-970, Brasil\\
}

\vspace{1cm}

\begin{abstract}
Motivated by the low-energy electron recoil spectrum observed by the dark matter experiment, XENON1T, at Gran Sasso laboratory, we interpret the observed signal not in terms of a dark matter particle, but rather in the context of a new light $Z^\prime$ gauge boson. We discuss how such a light $Z^\prime$ emerges in a Two Higgs Doublet Model augmented by an abelian gauge symmetry where neutrino masses and the flavor problem are addressed, in agreement with neutrino-electron scattering data.

\end{abstract}

\maketitle

\section{Introduction}

A collection of observations at dwarf galaxies, galaxy clusters, and the cosmic microwave background, and baryon acoustic oscillations have solidly confirmed the presence of a dark matter component in our universe that accounts for about 
 {\red 27\% } of the total energy budget. Its origin is still unknown, and one of the most compelling interpretations is elementary particles. In that regard, WIMPs (Weakly Interactions Massive Particles) and Axions are quite popular dark matter candidates (see \cite{Arcadi:2017kky} for a review). For this reason, several experiments have been searching them at colliders, direct, and indirect detection experiments. 

For a very long time, direct dark matter detection meant probing the dark matter-nucleon scattering, but as the detectors were upgraded and low background events better understood, electron recoils became interesting probes for light dark matter. The XENON1T detector is a prime example. The employed liquid-xenon time projection chamber was initially designed to detect WIMPs, but due to the large fiducial volume, low background rate, low threshold, such detectors are also powerful probes for light new particles, neutrino physics, and other dark matter candidates.

The dual phase liquid xenon TPC technology has leading sensitivities for dark matter masses above $6$~GeV \cite{Aprile:2018dbl}. In particular, the XENON10 \cite{Angle:2007uj}, XENON100 \cite{Aprile:2012nq}, and XENON1T \cite{Aprile:2017iyp} experiments have set stringent bounds on the dark matter-nucleon scattering cross-section, for isospin conserving dark matter interactions \cite{Yaguna:2019llp}. There are several important and complementary limits on the dark matter-nucleon scattering cross-section \cite{Akerib:2015rjg,Cui:2017nnn,Witte:2017qsy,Agnes:2018ves}, especially in the light dark matter mass region where other technologies yield tighter constraints \cite{Agnes:2018oej}.

In the exciting and competitive hunt for dark matter signals, the  XENON collaboration has recently reported a positive signal of  low-energy electronic recoil events. This corresponds to a release of data taken between February 2017 and February 2018 with the XENON1T detector, with an exposure of 0.65 tonne-years and low background rate of $(76\pm 2)~  {\rm events} ~ (tonne \times year \times keV)$ in the $1-30$~keV energy range \cite{Aprile:2020tmw}. An excess over known backgrounds was excitingly observed below $7$~keV, rising towards lower energies, peaking around 2-3 keV \cite{Aprile:2020tmw}. 


The XENON collaboration emphasizes that the excess might also be due to new backgrounds like tritium at a minute level, too small to be excluded for now, while the upcoming XENONnT detector will be able to differentiate. It is nevertheless tempting to study what kind of new physics might explain such an excess. This has already been the subject of several new physics interpretations, some dealing with new neutrino interactions \cite{Boehm:2020ltd,Amaral:2020tga,Bally:2020yid,AristizabalSierra:2020edu}, axions or Axion Like Particles (ALP) \cite{Takahashi:2020bpq} or in the context of dark matter physics \cite{Alonso-Alvarez:2020cdv,Fornal:2020npv,Kannike:2020agf,Su:2020zny,Harigaya:2020ckz,Du:2020ybt,Cao:2020bwd,Lee:2020wmh}. The majority of the interpretations are in tensions with astrophysics observable. For instance, as underlined by \cite{Boehm:2020ltd,AristizabalSierra:2020edu}, to interpret this excess by a new interaction between solar neutrinos and the electron one needs to introduce a light mediator ($\lesssim$ 0.1 MeV) which is already excluded by the physics of horizontal branch stars, and affect drastically the neutrino mean free path in supernovae. On the other hand \cite{DiLuzio:2020jjp} showed that the values of the coupling needed for a solar axion to fit with the XENON1T data is in conflict (up to $19 \sigma$ with stellar evolution.
Finally, models considering a dark matter origin to the electron recoil, usually necessitate a boosted dark matter ($v_{DM} \simeq 0.1$c \cite{Fornal:2020npv,Kannike:2020agf}), to reach the keV range of recoil energies\footnote{The electron being much lighter than the XENON nucleus, a typical galactic velocity $v \simeq 10^{-3} c$ is not enough to generate keV recoil energy.}.
Other authors employed multicomponent dark matter \cite{Bell:2020bes} where a heavier candidate scatter inelastically on the electron, producing a lighter state, or a hidden photon, product of a warm dark matter decay interacts with the electron \cite{Choi:2020udy}. 

In our work, we interpret the observed excess in terms of a complete model which leads to a light $Z^\prime$. It is a Two Higgs Doublet Model (2HDM) featuring an abelian gauge symmetry which naturally fits the XENON1T data. 2HDM models are popular extensions of the Standard Model (SM), but the general and original version suffers from flavor changing interactions and lacks neutrino masses. Via the introduction of an Abelian gauge symmetry, we can solve these two issues. In addition, a light $Z^\prime$ arises having the features needed to accommodate the XENON1T anomaly, in agreement with neutrino-electron constraints. 

Our work is structured as follows: In section II we present the model; in section III we discuss how we can fit the data; in section IV we introduce the relevant neutrino-electron constraints before summarizing our conclusions.

\label{sec_model}
\begin{table*}[!t]
\centering
\begin{tabular}{cccccccccc}
\hline 
Fields & $u_R$ & $d_R$ & $Q_L$ & $L_L$ & $e_R$ & $N_R$ & $\Phi _2$  & $\Phi_1$ & $\Phi_S$ \\ \hline 
Charges & $u$ & $d$ & $\frac{1}{2}(u+d)$ & $-\frac{3}{2}(u+d)$ & $-(2u+d)$ & $-(u+2d)$ & $\frac{1}{2}(u-d)$ & $\frac{1}{2}(5u+7d)$ & $2u+4d$ \\
$U(1)_{B-L}$ & $1/3$ & $1/3$ & $1/3$ & $-1$ & $-1$ & $-1$ & $0$ & $2$ & 2 \\
\hline
\end{tabular}
\caption{Anomaly free 2HDM-$U(1)_{B-L}$ model that explains neutrino masses via a type I seesaw, solves the flavor problem in 2HDM, and accommodates the XENON1T anomaly.}
\label{2HDMmodels}
\end{table*}

\section{The Model}

In the usual Two Higgs Doublet Model (2HDM), the scalar sector of the SM is extended 
by the addition of an extra Higgs doublet, besides the one already present. 
These scalars can be parametrized as,
\begin{equation}
\Phi _i = \begin{pmatrix} \phi ^+ _i \\ \left( v_i + \rho _i + i \eta _i \right)/ \sqrt{2}\end{pmatrix} , \text{\ \ \ \ \ \ } i = 1,2 .
\end{equation}
It is well known that general 2HDMs suffer from large Flavor Changing Neutral Interactions (FCNI) since the inclusion of the second doublet brings with it extra neutral scalars that mediate this kind of process at tree level. The stringent bounds from flavor physics \cite{Cogollo:2013mga,Lindner:2016bgg,Batell:2017kty,Sirunyan:2017uae,Krnjaic:2019rsv} severely constrain the corresponding Yukawa couplings, which demands a fine-tuning on these parameters.
%
%
This unnatural Yukawa suppression can be elegantly avoided by the introduction of an Abelian gauge symmetry $U(1)_X$ \cite{Ko:2012hd,Ko:2013zsa,Ko:2014uka,Ko:2015fxa,DelleRose:2017xil,Nomura:2017wxf,Botella:2018gzy}, 
as long as the two scalar doublets transform differently from each other under $U(1)_X$. This requirement is important in order to prevent the fermions from obtaining their masses from more than one source, which is the origin of FCNI in this kind of model \cite{Glashow:1976nt,Paschos:1976ay,Atwood:1996vj,Mahmoudi:2009zx,Crivellin:2013wna}.

One straightforward way to accomplish this is to make only one of the doublets responsible for the generation of all the SM fermion masses, which is always possible by a suitable choice of charges. Choosing the $\Phi _2$ doublet to play this role, we have, 
\begin{equation}
\label{2hdm_tipoI_u1}
- \mathcal{L} _Y = y_2 ^d \bar{Q} _L \Phi _2 d_R + y_2 ^u \bar{Q} _L \widetilde \Phi _2 u_R + y_2 ^e \bar{L} _L \Phi _2 e_R + h.c. .
\end{equation}

The many possible charge assignments for the fermions and scalars under this $U(1)_X$ symmetry are in general constrained by the demand of correct generation of fermion masses through the Yukawa interactions and also by the anomaly cancellation requirement, which has to be considered as the fermions are chiral under $U(1) _X$. In this model, the anomaly cancellation can be achieved even without the addition of extra fermions, although it is also possible to include them consistently if desired.

%
%

A massive $Z ^\prime$ gauge boson arises from the spontaneous breaking of the $U(1) _X$ symmetry. The $Z ^\prime$ mass and interactions are determined, to a large extent, by how this symmetry is broken. For instance, if a zero hypercharge scalar singlet is the only responsible for the $U(1) _X$ breaking, the $Z ^\prime$ mass becomes proportional to the VEV of this singlet. In this case, even if all the fermions are neutral under $U(1) _X$, they still interact with $Z ^\prime$ via kinetic mixing between $U(1) _X$ and the SM $U(1) _Y$, with couplings proportional to their electric charge but suppressed by the small kinetic mixing parameter, which is why $Z ^\prime$ is called a dark photon. In our 2HDM case, the $U(1) _X$ is broken by a scalar which transforms in a nontrivial representation of the SM gauge group, since at least one of the doublets is charged under $U(1) _X$, as required by the solution of the FCNI problem. Therefore, there will be mass mixing among $Z ^\prime$ and the SM gauge bosons. This mixing also induces interactions of $Z ^\prime$ with the fermions even if they are neutral under $U(1) _X$.

Nevertheless, whenever the fermions are charged under $U(1) _X$, the mass and kinetic mixing contributions become subdominant in face of the following interaction:
\begin{widetext}
\begin{equation}
\begin{split}
\mathcal{L} _{Z ^\prime} = & - \frac{1}{4} g_X \left[ \left( Q_{Xf} ^R + Q_{Xf} ^L \right) \bar{\psi} _f \gamma ^\mu \psi _f - \left( Q_{Xf} ^L - Q_{Xf} ^R \right) \bar{\psi} _f \gamma ^\mu \gamma _5 \psi _f \right] Z' _\mu,
\end{split}
\label{www1}
\end{equation}
\end{widetext}
where $Q^R_X$ ($Q^L_X$) are the left-handed (right-handed) $U(1)_X$ fermion charges and $g _X$ is the $U(1)_X$ coupling constant. 
This interaction is important in what follows, since it is responsible for the new contributions to the neutrino-electron scattering cross-section, as discussed in the next sections. 

Thus far, neutrinos are still massless. However, it has been shown that in this 2HDM-U(1) framework neutrino masses can be generated either by type I or type II seesaw mechanisms \cite{Campos:2017dgc,Camargo:2018uzw,Camargo:2019ukv,Cogollo:2019mbd}. In this work, we focus on the type I case, as described below.


\paragraph{Type I seesaw realization} 

The implementation of type I seesaw mechanism requires the introduction of three right-handed neutrinos $N _R$ to the particle content of the model. The neutrinos must be charged under $U(1) _X$ so that the Dirac term $y^{D} \bar{L} _L \widetilde \Phi _2 N_R$, which couples the right- and left-handed neutrinos, is allowed. 
The Majorana term, also required by the mechanism, becomes possible only if we include also a scalar singlet $\Phi_s = ( v _s + \rho _s + i \eta _s ) / \sqrt{2}$, which transforms as $\Phi _s \sim (1, 1, 0, q _{X s})$ under the group $SU(3) _c \times SU(2) _L \times U(1) _Y \times U(1) _X$. In this manner, we have the following Yukawa Lagrangian for the neutrinos,
\begin{equation}
\label{yukawa_neutrino_1}
- \mathcal{L} _\nu = y^{D} \bar{L} _L \widetilde \Phi _2 N_R + y ^R \overline{N _R ^c} \Phi _s N _R + h.c. .
\end{equation} 
With this Lagrangian, the type I seesaw mechanism is realized, yielding $ m _\nu = - m _D ^T M _R ^{-1} m _D $ for the mass of the active neutrinos and $ m _N = M _R $ for right-handed neutrinos, where $ m _D = y ^D v _2 / \sqrt{2} $ and $ M _R = \sqrt{2} y ^M v _s $, with $M _R \gg m _D$. As the active neutrino masses are inversely proportional to $v _s$, we assume the VEV hierarchy $v_s \gg v$, where $v = \sqrt{v_1^2+v_2^2}$.
%
%
%

Through Eqs.\ \eqref{2hdm_tipoI_u1} and \eqref{yukawa_neutrino_1} we successfully generate the masses to the neutrinos and other SM fermions after the spontaneous symmetry breaking. Note also that these equations enforce relations among the particle charges which comprise, together with the anomaly cancellation conditions, the set of constraints that the $U(1) _X$ charges must obey. Solving these constraints, we find that in general the charges of all fermions and scalars can be written in terms of $u$ and $d$, the charges of the right-handed up and down quarks, respectively, which remain independent. In this work, we focus on the $B-L$ solution, as shown in Table \ref{2HDMmodels}.

As for the $Z ^\prime$, once the scalar singlet, as well as the doublets, break the $U(1) _X$ symmetry, the $Z ^\prime$ mass gets contributions from both sources, 
\begin{equation}
\label{light_z_prime_mass}
m _{Z '} ^2 =  \frac{g _X ^2}{4} [ q _{Xs} ^2 v _s ^2 + ( Q _{X1} - Q _{X2} ) ^2 \frac{v _1 ^2 v _2 ^2}{v ^2} ] .
\end{equation}
As we assume $v _s \gg v$, the first term corresponding to the singlet contribution dominates. Notice that the coupling constant $g _X$ appears as an overall factor, so that for sufficiently small $g _X$, $Z ^\prime$ can be very light. \\

\section{Fit to Data}

In order to estimate the best-fit region in the $g_l g_\nu$ vs $m_{Z^\prime}$ plane we closely follow the procedure described in \cite{Boehm:2020ltd}. It relies on computing the energy spectrum which is given by,
\begin{equation}
\frac{dR}{dE_R}= F(E_R) \frac{A_T}{m_T} \int dE_\nu \frac{d\phi}{dE_\nu}\frac{d\sigma_\nu}{dE_R}   
\end{equation}

\noindent
where $E_R$ is the recoil energy, $A_T$ the exposure, $F(E_R)$ a function that account for the number of scattered electrons at a given energy, $d\phi/dE_\nu$ the spectra of solar neutrinos\footnote{For our analysis we used the primary fusion process in the Sun, also called the $pp$ process ($p+p \rightarrow$ $^2H + e^+ +\nu_e$) leading to the production of neutrinos up to $E_\nu \lesssim 400$ keV.} \cite{Cerdeno:2016sfi}. In the presence of light $Z^\prime$ gauge bosons, as the ones present in our 2HDM-$U(1)_X$ models, the change in the energy spectrum appears due to new neutrino-electron interactions mediated by the $Z^\prime$ model. We will consider  the $U(1)_{B-L}$ model, which contains the lagrangian,

\begin{equation}
\mathcal{L} \supset (\frac{g_\nu}{4} \bar{\nu}_L \gamma^\mu \nu_L + \frac{g_l}{4} \bar{l}\gamma^\mu l )Z^\prime_\mu.
\label{neutralEq}
\end{equation}

\noindent
We account for the change in the energy spectrum due to the presence of these new interactions via the differential cross-section which can be written

\begin{equation}
    \frac{d\sigma_\nu}{dE_R} = \frac{\sqrt{2}G_F \pi^{-1} m_e g_\nu\, g_{v}\, g_{e}}{16( 2E_R m_e+ m_{Z^\prime}^2)}+\frac{ m_e g_\nu^2 g_{e}^2 (2\pi)^{-1}}{256( 2E_R m_e+ m_{Z^\prime}^2)^2}
    \label{Eq:dsigma}
\end{equation}where $g_{v}=  2 \sin^2 \theta_W +1/2$ for the electron-neutrino and $g_{v}=  2 \sin^2 \theta_W -1/2$ for $\nu_\mu$ and $\nu_\tau$. Note that $g_e$ in Eq.\ref{Eq:dsigma} is the coupling appearing in Eq.\ref{neutralEq} which controls the strength of the $Z^\prime$ interaction with charged leptons.

\begin{figure}
    \centering
    \includegraphics[width=\columnwidth]{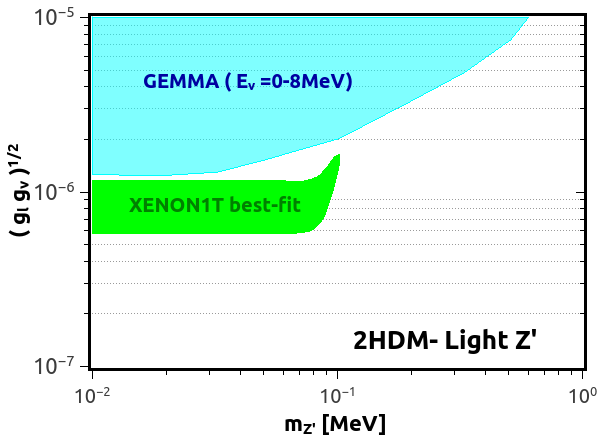}
    \caption{Best-fit region that accommodates the XENON1T anomaly overlaid with the existing limit from GEMMA. We highlight the inclusion of axial couplings yields no change in the best-fit region. The product $g_l g_\nu$ is the relevant in the analysis. One should note we use a different parametrization for the $Z^\prime$ lagrangian to properly compare our findings with others.}
    \label{fig1}
\end{figure}

Taking into account the neutral current shown in Eq.\ref{www1}, neglecting the kinetic mixing we have derived the region of parameter space that best fits the XENON1T anomaly. In our model, only vector currents are present, but we point out that the inclusion of axial-vector currents would not yield meaningful changes. Such axial-vector currents arise in different $U(1)_X$ symmetries as explored in \cite{Campos:2017dgc}. We highlight that we adopted a different parametrization for the neutral current involving the $Z^\prime$ gauge boson compared to previous works \cite{Boehm:2020ltd}, as we intended to follow the same notation of \cite{Campos:2017dgc}. Anyway, we plotted a green band in Fig.\ref{fig1}, which delimits the region of parameter space that accommodates the XENON1T anomaly at $1\sigma$ level. 

We notice that a typical feature will appear in the spectrum as long as the mediator $Z'$ is not decoupled. Indeed, it is the dependence in $2 m_e E_R$ in Eq.(\ref{Eq:dsigma}) that induces an enhancement in the number of events for low energies of recoil. This feature disappear in regimes where  
$\frac{m_{Z'}^2}{2 m_e} \gg E_R$. In other words, if one needs to fit with the XENON1T data, {\it i.e} a peak around $E_R=2.5$ keV, one needs to have $m_{Z'} \lesssim \sqrt{2 E_R m_e} \simeq 150$ keV, which is clearly visible in Fig.(\ref{fig1}).

As we are altering the neutrino-electron scattering rate, one should check whether this new interaction is also consistent with existing bounds from low energy probes of neutrino-electron scattering as we do in the next section.

\section{Neutrino-Electron Scattering Constraints}

 Our interpretation of the XENON1T excess in terms of a light $Z^\prime$ can be tested via precise measurements of the neutrino-electron scattering. The GEMMA experiment probed the neutrino magnetic moment using a HPGe detector near the Kalinin Nuclear Power Plant.
Having in mind that the SM neutrino interactions are suppressed with respect to its sensitivity in the $3-25$~keV energy range, only events in that energy gap were considered.  As GEMMA has a very low energy threshold, it is particularly sensitive to light gauge bosons with masses below the 100keV.

To derive GEMMA's bound we need to compute the event number in the
recoil energy bin $E_1 < E < E_2$,
\begin{equation}
    N= \int \phi (E_\nu) \sigma (E_\nu,E_1,E_2)dE_\nu,
\end{equation} where the flux is given by the collaboration and the cross section can be generally derived using the procedure found in \cite{Lindner:2018kjo} and then applied to our specific model.

For larger masses, the TEXONO \cite{Deniz:2009mu}, Borexino, LSND experiments offer better limits due to the interplay between precision and low energy threshold. For instance, we computed the TEXONO sensitivity to our model as well, but it resulted in $g_l g_\nu < 8 \times 10^{-5}$ for $m_{Z^\prime} < 100$~ keV. Similar or weaker bounds are found for the other experiments. Hence, as far as neutrino-electron scattering is concerned, GEMMA is the most relevant experiment. 

We highlight that in the derivation of the GEMMA limit was obtained using $\mu_\nu =3.2 \times 10^{-11}\mu_B$ \cite{Beda:2009kx,Beda:2010hk}. However, there are systematic uncertainties present which might weaken this limit. Moreover, we noted that the inclusion of axial-vector couplings yields no visible change in the best-fit region. 

It is clear that the next generation of direct detection dark matter experiments, namely soon XENONnT \cite{Aprile:2015uzo} and LZ \cite{Mount:2017qzi}, somewhat later Darkside \cite{Aalseth:2017fik}, will have much larger exposures which will allow to further study the excess. Signs of annual modulation, would for example point to new physics, while this is not expected for background explanations. With sufficient statistics it should also become possible to distingish if the excess goes like $1/T$ as expected for a large magnetic moment or like $1/T^2$ like in our model. The fact that Darkside uses a different target will help further to study a signal of new physics and to reject explanations with unknown backgrounds.


\section{Conclusions}

We entered an exciting era where dark matter detectors can probe neutrino physics. Tonne-size instruments begin to be more sensitive than neutrino dedicated experiments themselves. We discussed a light $Z^\prime$ model motivated by the low-energy electron recoil spectrum measured by the XENON1T experiment, which reached unprecedented sensitivity. The signal challenges the typical dark matter hypothesis and seems to beg for an alternative explanation. We interpreted the observed signal, not in terms of a dark matter particle that interacts with electrons, but rather with a light $Z^\prime$ embedded in a well-motivated model, based on Two Higgs Doublets and an Abelian gauge symmetry. We have shown that our model is capable of solving the flavor problem in Two Higgs Doublet Models, explain neutrino masses via a type I seesaw mechanism, and address the XENON1T anomaly in agreement with the stringent bound from the GEMMA experiment on the neutrino-electron interaction at low energies.

\section{Note}
During the completion of our work, we noticed that \cite{AristizabalSierra:2020edu} and \cite{Khan:2020vaf} reached the same conclusion with models including also a light mediator.

\section*{Acknowledgments}
The authors thank Pedro Machado por discussions. TM and FSQ thanks UFRN and MEC for the financial support. FSQ also acknowledges the CNPq grants 303817/2018-6 and 421952/2018-0, and the ICTP-SAIFR FAPESP financial support grant 2016/01343-7. This work was supported by the Serrapilheira Institute (grant number Serra-1912-31613) and by the France-US PICS MicroDark.
This project has received funding/support from the European Unions Horizon 2020 research and
innovation programme under the Marie Skodowska-Curie grant agreements Elusives ITN No. 674896
and InvisiblesPlus RISE No. 690575. We thank the High Performance Computing Center (NPAD) at UFRN for providing computational resources.
\newpage

\bibliographystyle{reffixed}
\bibliography{darkmatter}

\end{document}